\documentclass[aps,prl,twocolumn,superscriptaddress,longbibliography,10pt]{revtex4-2}

\usepackage{graphicx}
\usepackage{amsmath,amssymb}
\usepackage{bm}
\usepackage{mathrsfs}
\usepackage[T1]{fontenc} 
\usepackage{lmodern}
\usepackage[dvipsnames]{xcolor}
\usepackage{marvosym}
\usepackage{hyperref}
\hypersetup{colorlinks = true,citecolor = blue,urlcolor = black}

\usepackage{url}

\newcommand{\Peclet}{P\'eclet}

\graphicspath{{./Figures/}{./model}}

\begin{document}

\title{Topological transition in filamentous cyanobacteria: from motion to structure}
\author{Jan Cammann}
\affiliation{Interdisciplinary Centre for Mathematical Modelling and Department of Mathematical Sciences, Loughborough University, Loughborough, Leicestershire LE11 3TU, United Kingdom}
\author{Mixon K. Faluweki}
\affiliation{School of Science and Technology, Nottingham Trent University, Nottingham NG11 8NS, UK}
\affiliation{Malawi Institute of Technology, Malawi University of Science and Technology, S150 Road, Thyolo 310105, Malawi}
\author{Nayara Dambacher}
\affiliation{School of Science and Technology, Nottingham Trent University, Nottingham NG11 8NS, UK}
\affiliation{School of Biosciences, University of Nottingham, Sutton Bonington Campus, LE12 5RD, UK}
\author{Lucas Goehring}
\email{E-mail: lucas.goehring@ntu.ac.uk}
\affiliation{School of Science and Technology, Nottingham Trent University, Nottingham NG11 8NS, UK}
\author{Marco G. Mazza}
\email{E-mail: m.g.mazza@lboro.ac.uk}
\affiliation{Interdisciplinary Centre for Mathematical Modelling and Department of Mathematical Sciences, Loughborough University, Loughborough, Leicestershire LE11 3TU, United Kingdom}

\begin{abstract}
Many active systems are capable of forming intriguing patterns at scales significantly larger than the size of their individual constituents. 
Cyanobacteria are one of the most ancient and important phyla of organisms that has allowed the evolution of more complex life forms. Despite its importance, the role of motility on the pattern formation of their colonies is not understood. 
Here, we investigate the large-scale collective effects and rich dynamics of gliding filamentous cyanobacteria colonies, while still retaining information about the individual constituents' dynamics and their interactions. 
We investigate both the colony's transient and steady-state dynamics and find good agreement with experiments. We furthermore show that the \Peclet{} number and aligning interaction strength govern the system's topological transition from an isotropic distribution to a state of large-scale reticulate patterns. Although the system is topologically non-trivial, the parallel and perpendicular pair correlation functions provide structural information about the colony, and thus can be used to extract information about the early stages of biofilm formation.  
Finally, we find that the effects of the filaments' length cannot be reduced to a system of interacting points. 
Our model proves to reproduce both cyanobacteria colonies and systems of biofilaments where curvature is transported by motility.
\end{abstract}

\maketitle

\section{Introduction}

\begin{figure}[ht]
    \centering
    \includegraphics[width=\columnwidth]{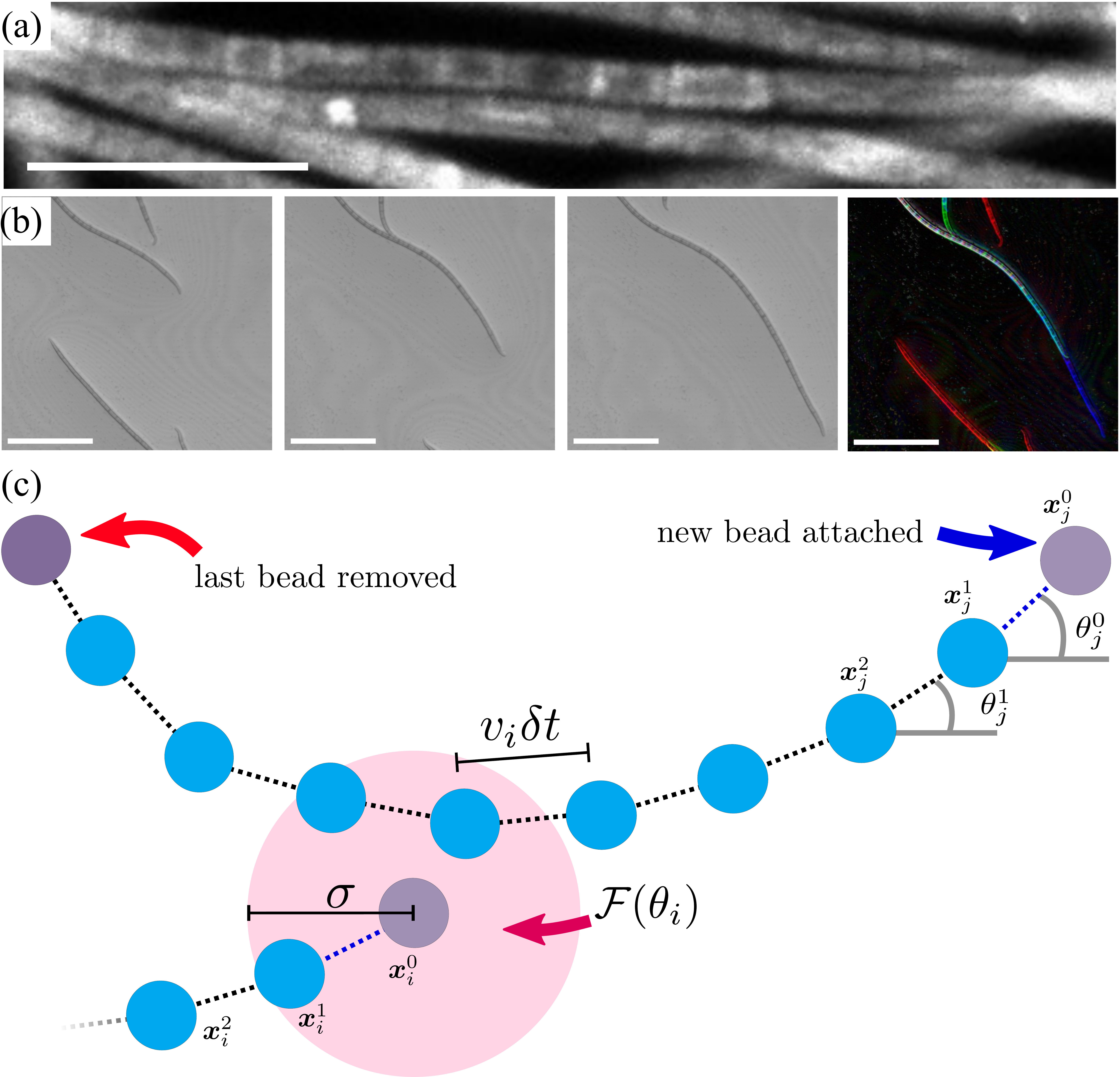}
    \caption{Dynamics of gliding motility. (a) Close-up micrograph of \textit{Oscillatoria lutea} filaments, composed of chains of individual cells. The scale bar is $30~\mu$m.
    (b) A time lapse of three micrographs demonstrates filament motion, by which each filament's body glides along the track laid down by its head; the interval between frames is 51 s; the scale bars in each frame are $100~\mu$m. The rightmost image highlights this path-tracking dynamics by overlaying the three images using shadings of different colors. 
    (c) Schematic depiction of the model filaments, as discretized into chains of connected beads. Beads are placed at a fixed distance $v_i\delta t$ apart. Upon advancing time by an increment $\delta t$, the last bead of a filament is removed and placed as its new head.  Non-reciprocal interactions occur when a filament's head (filament $i$) is within a distance $\sigma$ (pink disk; not drawn to scale) from any bead of another filament, $j$. The head experiences a nematic aligning interaction $\mathcal{F}(\theta_i)$ while filament $j$ remains unaffected by this interaction.}
    \label{fig:model_discretisation}
\end{figure}

Living matter is fundamentally characterized by its ability to self-organize \cite{maturana1975organization,kauffman1991antichaos,kauffman1993origins,karsenti2008self}. This is visible at length scales spanning a staggering range of about nine orders of magnitude, from the few microns of the complex structures present inside a eukaryotic cell, such as its nucleus, the endoplasmic reticulum, or the length of a microtubule 
\cite{yates2005proteomics,shibata2009mechanisms,shelley2016dynamics}, to the complex spatial organization in bacterial biofilms extending over millimeters \cite{stoodley2002biofilms,mazza2016physics,sauer2022biofilm}, to the scale of flocks of birds \cite{bialek2012statistical,cavagna2014bird} or schools of fish \cite{katz2011inferring}, up to kilometer-long herds of wildebeest~\cite{gueron1993self,couzin2003self}. Starting from the seminal work of Vicsek \textit{et al.} \cite{vicsek1995}, the self-organization of large collectives of simple objects has garnered considerable interest within the field of active matter~\cite{vicsek1995,toner1995,simha02,ramaswami2010,marchetti2013,elgeti2015,fruchart2021, Shi2018, needleman2017,doostmohammadi2018}. Active systems are defined by the injection of energy into individual elements, which is converted into directed motion \cite{bechinger2016active}. This energy injection explicitly breaks time-reversal symmetry \cite{bowick2022symmetry} and allows active matter systems to exhibit  emergent properties not observable in equilibrium systems \cite{marchetti2013,elgeti2015}.

A simplifying assumption involved in the modeling of many active-matter systems is to treat their basic components as point-like particles~\cite{vicsek1995,toner1995,ramaswami2010} or stiff rods \cite{bar2020, ginelli2010, baskaran2008}, each with its own orientation, and a few rules for their motion and pairwise interactions. Archetypes of these systems are the Vicsek model \cite{vicsek1995,vicsek2012collective,martin2024fluctuation}, the active Brownian particle model \cite{romanczuk2012active,solon2015active,wagner2017steady}, where self-propelled particles interact with neighbors via spherically-symmetric steric repulsion,  and the active nematic model \cite{sanchez2012,giomi2015geometry}, where filaments are described with a single orientation, in analogy to liquid crystal molecules. 

In contrast, long, flexible filaments such as microtubules or active polymers are typically described by an orientation $\theta$ that varies along their arc length $s$.  Their dynamics are then commonly modeled using techniques adopted from polymer physics   \cite{liverpool2001, isele2015, jiang2014, duman2018, bianco2018, anand2018, saintillan2018extensile,joshi2019,martin2019,fily2020,vliegenthart2020, winkler2020, peterson2020,deblais2020phase,du2022,abbaspour2021, struebing2020}, where filaments are represented by discrete monomers connected together through bonding potentials, allowing each link within the chain to have its own laws of motion.

Cyanobacteria are one of the most important~\cite{blankenship2010early,cardona2019early,sanchez2022cyanobacteria} and ancient~\cite{mayall1981algal,wright1981organism,bivzic2020,Sumner1997} groups of species to have evolved on Earth, with evidence of their presence in the fossil record dating back to at least two billion years ago \cite{demoulin2019cyanobacteria,schirrmeister2013evolution}. Their ancestors first evolved oxygenic photosynthesis \cite{sanchez2022cyanobacteria}, which changed the chemical composition of the atmosphere, thus paving the way for the evolution of animals. In green plants and eukaryotic algae photosynthesis is carried out in the plastids, organelles now recognized to be endosymbiotic cyanobacteria,  one of the two major examples of endosymbiosis in the history of life 
\cite{archibald2015endosymbiosis} (the other is the case of mitochondria). 
In terms of sheer number of cells, cyanobacteria are also the most common type of life that has ever lived \cite{crockford2023geologic}. They are 
found in extreme environments like hot springs, deserts and the polar regions \cite{whitton2012introduction}, but also in everyday streams and fountains. 
A particularly diverse prokaryotic phylum, species of cyanobacteria range from unicellular to multicellular filamentous and branching forms \cite{schirrmeister2013evolution,Schirrmeister2011}. 

Filamentous cyanobacteria grow as long hair-like chains of cells called trichomes, see Fig.~\ref{fig:model_discretisation}(a), and are a particularly important type of microorganism; for example, they dominate the ecology of shallow lakes \cite{scheffer1997dominance}, and due to their size and structure they can be used in wastewater management more efficiently than unicellular microbes \cite{markou2011cultivation}.
Numerous species of filamentous cyanobacteria exhibit gliding motility, which is instrumental in the formation of biomats \cite{hoiczyk2000,Shepard2010UndirectedMats}. 
 
These gliding cyanobacteria are becoming recognized as the archetype of a distinct class of active filaments~\cite{faluweki2023,kurjahn2024, Kurjahn2022,repula2024,gong2023}. Unlike active polymers, their motion cannot be represented as resulting from active forces tangential to monomer-like elements \cite{winkler2020,franks2016}.  Gliding cyanobacteria exert negligible forces on each other, as opposed to the stresses that they apply on their substrate; thus, physical
interactions are different than active nematics based on microtubule-kinesin or actin-myosin \cite{schaller2010}.
The physical properties that set gliding cyanobacteria apart from other active filament systems, and which help define their own class of active matter, active spaghetti, include: (\textit{i}) a large aspect ratio, (\textit{ii}) gliding motility induced by polar forces, (\textit{iii}) path-tracking dynamics, where the body of the filament follows the track laid down by its head (also known as metameric locomotion~\cite{du2022}), and (\textit{iv}) non-reciprocal aligning interactions.

Cyanobacteria can form biomats \cite{del2018biofilm}, and their elongated species can form reticulate biomats \cite{cuadrado2018,Shepard2010UndirectedMats}.
Their reticulate features are irregular network-like patterns emerging from the self-organization of trichomes first into dense bundles and then into a network topology.  These patterns contribute to rapid collective responses~\cite{Castenholz1968,Pfreundt2023}, mechanical properties~\cite{Shepard2010UndirectedMats} and act as a template for more complex 3D structures~\cite{Shepard2010UndirectedMats,Mackey2017}. 
We recently studied the reticulate network that emerges as a self-organized pattern of gliding filamentous cyanobacteria~\cite{faluweki2023}. The characteristic scale of this network is determined by a balance of activity and fluctuations \cite{faluweki2023}. 

Here, we explore the dynamics, structure, and phase behavior of a system of active filaments, modeled after the behavior of cyanobacteria. 
We investigate the large-scale collective effects  of cyanobacteria colonies, while still retaining information about the individual constituents' dynamics and their interactions. 
Combining numerical simulations with theoretical arguments, we uncover a rich dynamics for this model of gliding filaments as their \Peclet{} number $\mathrm{Pe}$, area coverage $\Phi$, and  filament length $L$ are varied.  As it develops towards a statistically steady state, this model shows transient behaviors that closely resemble corresponding experiments with filamentous cyanobacteria.  We also find that the filaments' effective translational diffusivity grows with the \Peclet{} number, $\mathrm{Pe}$. 
The nonequilibrium phase diagram of the system exhibits a transition from isotropic to reticulate patterns as the \Peclet{} number and area coverage, or density, increase. Finally, inspired by polymer physics, we compute the parallel and perpendicular pair correlation functions for the filaments. The features of these correlation functions carries information about the structure of the reticulate network and faithfully predicts its characteristic length scale. 

Taken together, our results identify the curvature fluctuations and aligning interactions as the mechanisms underpinning the nonequilibrium transition to the reticulate pattern in filamentous cyanobacteria.

\section{Methods}
\subsection{Model for gliding cyanobacteria}

\begin{figure*}
    \centering
    \includegraphics[width=\textwidth]{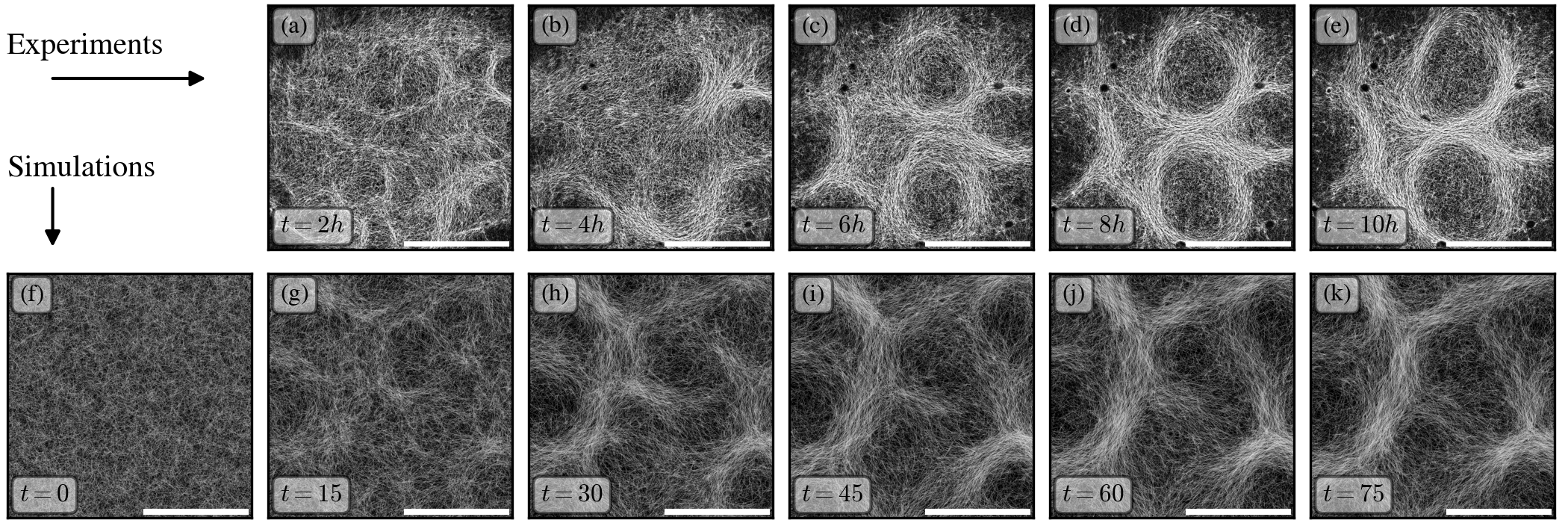}
    \caption{Time evolution of systems of filaments forming into reticulate structures. 
    (a)-(e) Micrographs of experiments performed with \textit{Oscilatoria~lutea}, where $v_0=3\,\mu$m/s and $\tau=8\,$min \cite{faluweki2023}.  
    (f)-(k) Simulations of a colony initialized with randomly oriented filaments, using a filament length $L=1$, 
    a \Peclet{} number $\mathrm{Pe}=3.5$ and an area coverage $\Phi=1$. The field of view shown and time between snapshots match the  experimental measurements of \textit{O.~lutea} (i.e., simulated time $t/\tau = 15$ corresponds to 2 h). The scale bars in the experimental images are $10\,$mm and scale bars in simulations $10\,$mm$/(v_0\tau)$.  In both experiments and simulations the initial pattern forms with smaller features that coarsen slightly to achieve a stable length scale over a period of a few hours ($t\approx 50$). 
    }
    \label{fig:progress}
\end{figure*}

Empirical observations guide the development of our model of the gliding motion of cyanobacteria, which then predicts the emergent behavior of cyanobacteria colonies.  
When gliding on surfaces, the head of a trichome lays out a path that the rest of the filament follows very closely; this is an analogue of the metameric motion observed in segmented animals, such as annelids and myriapods \cite{du2022}. 

Figure~\ref{fig:model_discretisation}(b) shows a time-series of micrographs of \textit{Oscillatoria lutea}, which are then overlaid to demonstrate how the body of each filament follows the track of its head. 
To capture this behavior, we discretize the trichomes as chains of connected beads. In contrast to active polymer models, however, these beads do not have active forces applied tangent to the length of the filament, nor is there an independent random force (e.g.\ thermal noise) on each bead, nor any reaction force.  Rather, the only forces relevant to the motion of a filament are those applied to its head.
Filaments can glide over one another with no discernible friction, as such we do not include any steric repulsion in our quasi-2-dimensional (2D) model.

A filament moving at a constant gliding speed $v_i$ is modeled as a string of beads with a fixed spacing $v_i\delta t$, where $\delta t$ is the time step of the numerical integration. 
The gliding speed $v_i$ of each filament is constant in time, and drawn from a normal distribution with average $v_{0}$ and standard deviation $\delta v$; this is in agreement with experimental observations \cite{faluweki2023}, and avoids spurious synchronization effects. 
Within any one simulation, all filaments have the same length $L$. Therefore, the number of beads in each filament is $L/(v_i \delta t)$. Because $v_i$ is normally distributed, different filaments will have a slightly different number of beads. 

To advance the filament by one time step, we remove the last bead in the chain (the tail), and add a new bead as the new head. 
A similar approach has been taken to study the dynamics of isolated filaments \cite{du2022}, and of filaments on a lattice \cite{schaller2010}. 
In this way, for each bead $\alpha$ on any filament $i$ the resulting path-tracking motion for the beads' positions $\mathbf{x}_{i}^\alpha$ can be expressed as 
\begin{align}
    \mathbf{x}_{i}^\alpha(t)=\mathbf{x}_{i}^{\alpha-1}(t-\delta t)\,\text{,}
\end{align}
for all beads except the filament's head, $\alpha = 0$, as sketched in Fig.~\ref{fig:model_discretisation}(c).

The head bead ($\alpha=0$) is where interactions take place, and so is treated separately. 
Dropping the index $\alpha$, a filament's head moves forward at a fixed speed,
\begin{equation}
    \frac{d\mathbf{x}_i}{dt}=v_i \mathbf{\hat{t}}_i.
\end{equation}
Here, $\mathbf{\hat{t}}_i\equiv (\cos\theta_i,\sin\theta_i)^\mathrm{T}$ 
is the unit vector tangent to the filament at the head's position, and defines the head's orientation $\theta_i$. 

The gliding motion of a cyanobacteria filament is characterized by its orientation and curvature. The curvature is mathematically well defined and has an empirically measurable exponential autocorrelation function with autocorrelation time $\tau$.  
A Wiener process acting on the angle $\theta$ does not satisfy these properties, that is, the curvature is not a well-defined quantity, and the autocorrelation of the curvature does not exist (and hence cannot be an exponential function).  The simplest model that captures these properties is an Ornstein--Uhlenbeck process describing the curvature diffusion of the filament paths~\cite{faluweki2023}.
In this case, the angular speed $\omega_i$ and orientation $\theta_i$ of the head of filament $i$ evolve as
\begin{align}
    \frac{d\omega_i}{d t}= & -\frac{1}{\tau}\frac{d\theta_i}{dt}+\sqrt{2D_\omega}\xi(t)\label{eq:dt_omega}\\
    \frac{d\theta_i}{dt}= & ~ \omega_i-J\mathcal{F}(\theta_i), \label{eq:dt_theta}
\end{align}
where $\tau$ is the curvature autocorrelation time, $D_\omega$ is a diffusion coefficient quantifying the strength of the fluctuations in curvature, and $\xi_i(t)$ is a Gaussian white noise with zero mean and unit variance, $\langle \xi(t)  \rangle=0$, $\langle \xi(t) \xi(t') \rangle=\delta(t-t')$.

\begin{figure}
    \centering
    \includegraphics[width=1.\linewidth]{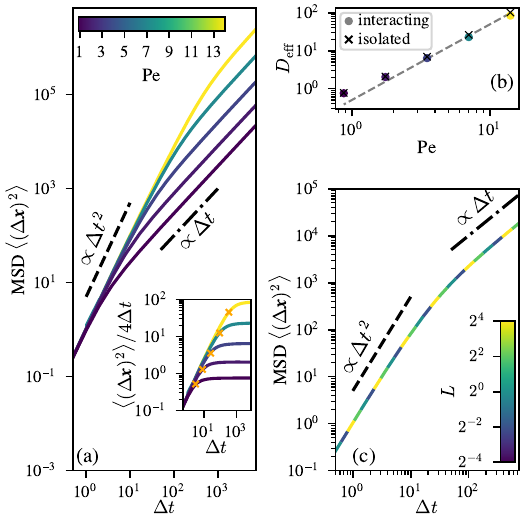}
    \caption{Dynamics of filaments in simulations. 
    (a) The mean square displacement (MSD) of the filaments' heads exhibits clear ballistic and diffusive regimes, as shown here for different $\mathrm{Pe}$ at fixed $\Phi = 1$ and $L=1$. The inset shows how the MSD converges to a long-time diffusive behavior, $\langle(\Delta x)^2\rangle \rightarrow 4D_\text{eff}\Delta t$. The orange crosses show the crossover time $t_\times=4D_\text{eff}/v_0^2$ between ballistic and diffusive regime.  This implies that filaments meander throughout the network, rather than being trapped within specific looped paths. 
    (b) The effective translational diffusion coefficient increases with \Peclet{} number for the fully interacting system. Switching off the interactions ($g=0$ in Eq.~\eqref{eq:dthetadt_nondim}) increases slightly $D_\mathrm{eff}$, with the deviation growing with $\mathrm{Pe}$. The dashed line represents the relation $D_\mathrm{eff}= \frac{1}{2}v_0^2\tau \mathrm{Pe}^2$.
    (c) Filament length has no discernible effect on the MSD, as shown here for the case of $\mathrm{Pe}=3.5$.}
    \label{fig:msd}
\end{figure}

\begin{figure}[t]
    \centering
    \includegraphics[width=\linewidth]{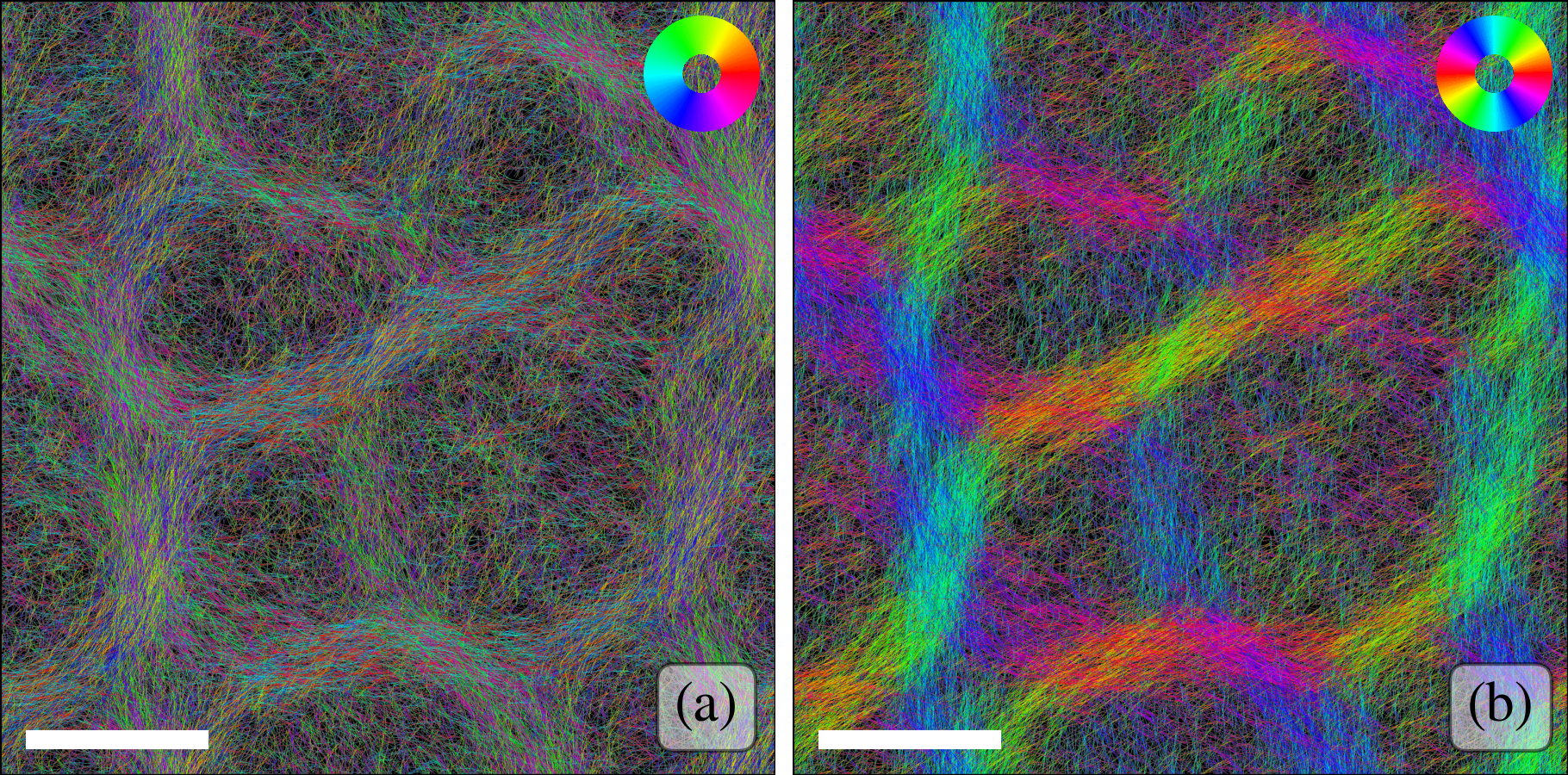}
    \caption{The reticulate bundles are nematic in nature. The two panels show the same reticulate pattern of a simulation with $\mathrm{Pe}=3.5$, $\Phi=1$ and $L=1$ at steady state, with filaments colored according to their local tangent (orientation) using (a) polar and (b) nematic mappings. The scale bars are $10\,$mm$/(v_0\tau)$. For the polar case, no preferential direction is discernible within the filament bundles. In the nematic mapping, where orientations $\theta$ and $\theta+\pi$ are assigned the same color, the bundles appear in bright colors, highlighting their nematic structure.}
    \label{fig:nematic-angles}
\end{figure}

The last term in Eq.~\eqref{eq:dt_theta} describes the interactions, of strength $J$, between nearby filaments, where $\mathcal{F}(\theta_i)=N_{ij}^{-1}\frac{\partial}{\partial \theta_i}\sum_{i\sim j}\cos(\theta_i-\theta_{j}^{\bar{\alpha}})$ is a Lebwohl--Lasher interaction \cite{lebwohl1972nematic,breier2018emergence} inducing nematic alignment between the filaments, $N_{ij}$ is the number of filaments within an interaction distance $\sigma$ of the \mbox{$i$-th} filament's head, $\sum_{i\sim j}$ denotes the sum over these filaments, and the orientation angle $\theta_{j}^{\bar{\alpha}}$ refers to the bead closest to the \mbox{$i$-th} head $\mathbf{x}_i$. The interaction has a short, finite range $\sigma$, which is consistent with the filaments' effective diameter; when the distance between the head of  filament $i$ and any bead of filament $j$ is less than $\sigma$, the orientation of the head of filament $i$ undergoes a deflection due to the Lebwohl--Lasher interaction, however, the bead of filament $j$ remains unperturbed. This makes the interactions fundamentally non-reciprocal, and it is predicated on the experimental observation that when two trichomes meet, only the incident filament's head participates in the alignment process, while the other filament is not deflected \cite{faluweki2023}. Non-reciprocal interactions explicitly break detailed balance \cite{loos2020irreversibility,knevzevic2022collective} and are common in living active matter \cite{nagy2010hierarchical}.
A schematic depiction of the non-reciprocal nematic interactions used in our model is shown in Fig.~\ref{fig:model_discretisation}(c).

Without any filament-filament interactions, Eqs.~\eqref{eq:dt_omega}-\eqref{eq:dt_theta}
produce a normal distribution of angular velocities with zero mean and variance $\langle \omega^2\rangle = D_\omega \tau$. For noninteracting filaments moving at an average speed $v_0$ this translates into a distribution of filament curvatures with zero mean and standard deviation $\delta \kappa = \sqrt{\langle \omega^2\rangle}/v_0 = \sqrt{D_\omega \tau}/v_0$.

Given the equations of motion \eqref{eq:dt_omega} and \eqref{eq:dt_theta}, a P\'eclet number may be defined as $\mathrm{Pe}=v_0/(l\sqrt{D_\omega\tau})$ with a length scale $l$.  This dimensionless number may be understood as a ratio of how quickly curvature $\kappa$ is transported within the system, compared with the time it takes to randomize a filament's curvature.

\begin{figure*}
    \centering
    \includegraphics{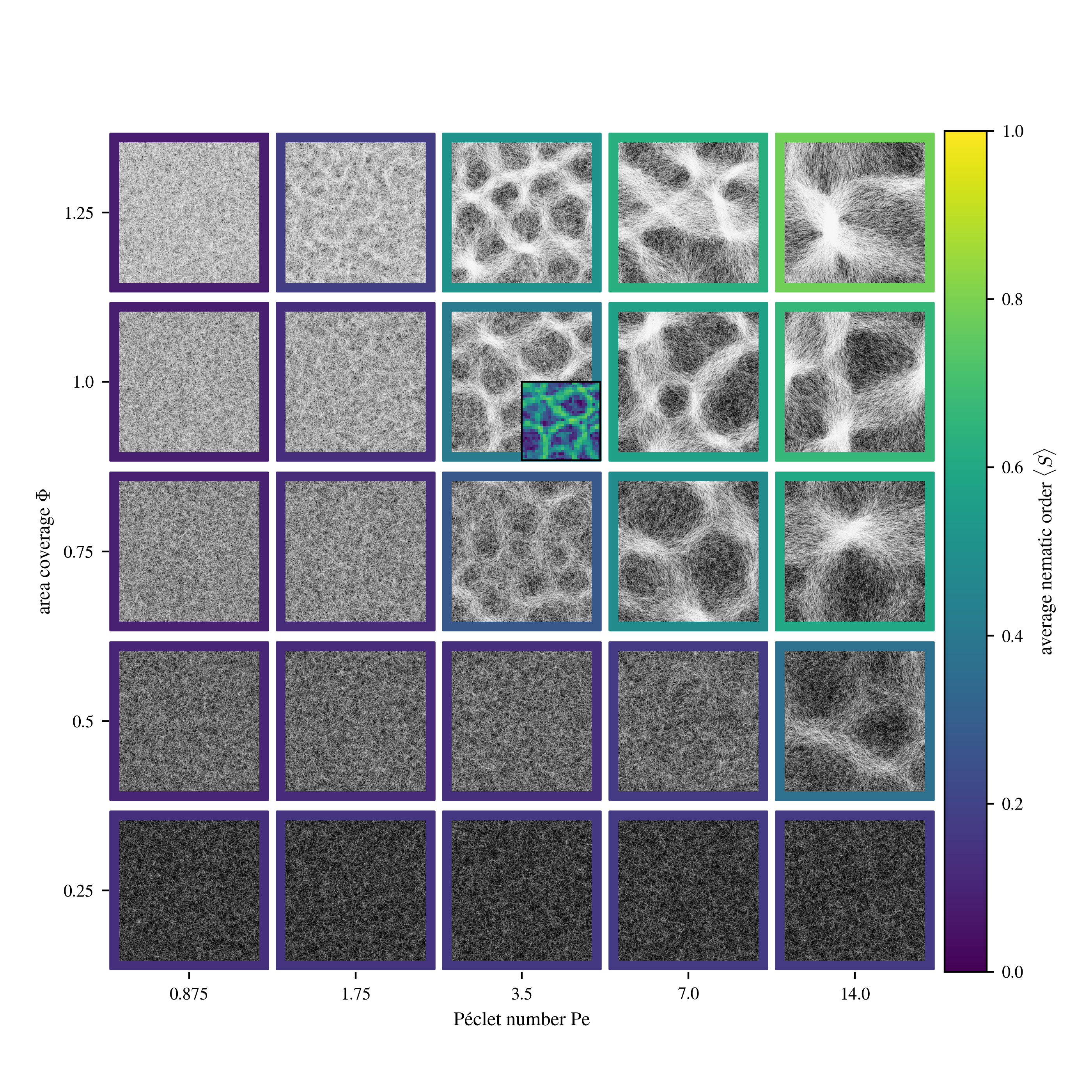}
    \caption{Emergence of reticulate patterns. The mosaic plot shows the steady-state behavior of the system at different area coverage $\Phi=0.25, 0.5, 0.75, 1.0, 1.25$ and \Peclet{} numbers $\mathrm{Pe}=0.875, 1.75, 3.5, 7.0, 14.0$. The individual panels show simulation snapshots at steady state in a $30\times30$ domain with periodic boundary conditions. Filament length is kept constant at $L=1$. The panels' frames indicate the system's average nematic order $\langle S \rangle$, calculated as an average of $1\times 1$ sized blocks. With increasing $\mathrm{Pe}$, the system transitions from a disordered state into one showing reticulate patterns of nematic bundles. The inset at $(\mathrm{Pe},\Phi)=(3.5, 1.0)$  shows an example of the spatially resolved nematic order parameter $S_\text{loc}$; regions of higher density correlate with higher local nematic order. Bundles show a high degree of ordering, whereas dilute  regions in between are less ordered.
    }
    \label{fig:combined_pe_rho}
\end{figure*}

To explore the general state diagram of our system, we nondimensionalize the equations of motion by setting $t=t_0 \tilde{t}$, $\mathbf{x}=l \tilde{\mathbf{x}}$, and by choosing $t_0=\tau$, the autocorrelation time, as the characteristic unit of time,  and  $l=v_0\tau$ as the characteristic length. The nondimensional equations of motion read
\begin{align}
    \frac{d\tilde{\mathbf{x}}}{d\tilde{t}}= & ~ \tilde{v}_i \mathbf{\hat{t}}_i \label{eq:dxdt_nondim}\\
         \frac{d\tilde{\omega}_i}{d \tilde{t}}= & -\frac{d\theta_i}{d\tilde{t}}+\frac{\sqrt{2}}{\mathrm{Pe}} \tilde{\xi}(\tilde{t}) \label{eq:domegadt_nondim}\\
    \frac{d\theta_i}{d\tilde{t}}= & ~ \tilde{\omega}_i-g\mathcal{F}(\theta_i) \label{eq:dthetadt_nondim}
\end{align}
where $\tilde{v}_i=v_i/v_0$, the dimensionless interaction rate $g=J\tau$, and the  \Peclet{} number naturally arises as $\mathrm{Pe}=1/\sqrt{D_\omega \tau^3}$.
In dimensionless units the width of the curvature distribution is given by the inverse P\'eclet number $\mathrm{Pe}^{-1}$, such that filaments with high $\mathrm{Pe}$ are on average straighter.
In the following, we will drop the tilde signs for ease of notation.

\subsection{Simulations}
Equations \eqref{eq:domegadt_nondim} and \eqref{eq:dthetadt_nondim} are discretized in time using the Euler--Maruyama method with a timestep of $\delta t=10^{-3}$, while equation \eqref{eq:dxdt_nondim} is discretized as discussed above and shown in Fig.~\ref{fig:model_discretisation}. 
We perform large-scale simulation of up to $N_t=1\,440\,000$ trichomes in a 2D domain of area $A = 30\times 30$,
for varying area coverage $\Phi =  \rho L (2\sigma)$ where $\rho=N_t/A$. We employ periodic boundary conditions.
We fix the width of the self-propulsion speed distribution as 
$\delta v=0.25$.
We choose nondimensional values $\sigma = 0.005$ and $g=2$, corresponding to a filament width of $d=\sigma v_0 \tau \approx 7\,\mu$m and an interaction rate of $J=g/\tau\approx 0.004\,$s$^{-1}$, close to the values measured experimentally for \textit{O. lutea} in \cite{faluweki2023}, up to some numerical rounding for the sake of simplicity.
Given our choices of parameters, a filament of length $L=1$ and self-propulsion speed $\tilde{v}_i=1$ is discretized with $1000$ beads. 
The choice of timestep ensures that $\tilde{v}_i \delta t< \sigma$ such that that every interaction between filaments is captured.
For simplicity, filaments are initially straight, but they rapidly (i.e.\ over timescales $t \approx 1$) develop curvature due to noise and alignment effects.
Note that area coverage $\Phi>1$ is possible, as there are no excluded volume effects.  In our simulations, we explore a range of $0.25\leq\Phi\leq1.25$ by changing the number of trichomes, and a range of $2^{-4}\leq L \leq 2^4$ by changing the number of beads per filament, while keeping $\sigma$, $\delta t$, $\delta v$, and $g$ fixed.

\subsection{Experiments}

\textit{Oscillatoria lutea} (SAG 1459-3) was cultivated in a medium of BG11 broth diluted 1:100 with deionised water, at $20 \pm 1^{\circ}$C with a  $10 \pm 2$ $\mu$mol m$^{-2}$ s$^{-1}$ photon flux on a 16\,h day + 8\,h night cycle, following Ref.~\cite{faluweki2023}.  For imaging, cultures were dispersed by gentle agitation and added dropwise into a 6-well plate (34 mm well diameter), three-quarters filled with medium.  Tile-scans of wells were collected using a confocal laser scanning microscope (Leica TCS SP5) in fluorescence mode (514 nm excitation, 620--780 nm observation).  Images were processed using ImageJ and MATLAB.

The dynamical properties of this strain were recently measured~\cite{faluweki2023}. We briefly recapitulate the relevant details here, and then recast these into dimensionless terms.   The filaments appear under the microscope as flexible cylinders with lengths $L = 1.5 \pm 0.5$ mm (mean $\pm$ standard deviation) and cross-sectional radius $r = 2.1 \pm 0.1 \,\mu$m.  Time lapse images of filament motion were taken using confocal imaging, with 512$\times$512 pixel (1.55$\times$1.55 mm$^2$) frames at a scan rate of 400 Hz. 
Individual filament behavior was measured by tracking 23 isolated filaments over up to three hours each, and applying thresholding and skeletonization.  The filaments glide at $v_0 =  3.0 \pm 0.7\,\mu$m/s, as measured by the motion of the midpoints of filament skeletons.  They move on smoothly curving trajectories, with a local curvature that fluctuates with time.  Curvature autocorrelation times of $\tau = 470 \pm 270$ s were measured using the filament midpoint tracks, and are consistent with measurements made by fitting circular arcs to the filament skeletons.  
Interacting cyanobacteria were observed at filament densities of between $\rho\approx 2$--$100$ mm$^{-2}$, in 6-well plates prepared as described above, but after 72 hours of incubation to ensure measurement of a dynamical steady state.  The filaments interact with each other when their paths cross.  This contact interaction appears stochastic: the intersecting filament either turns to follow alongside its neighbor, or glides over/under it without deflecting.  Nematic alignment results, with aligning responses seen in 16 of 400 tracked crossings, mostly involving crossings at angles below about $20^\circ$.  Distributions of filament curvature were measured  by fitting circular arcs to filament skeletons. At densities of $\rho =$ 30 mm$^{-2}$ and above these distributions were consistent with a mean of zero, and a standard deviation of $\delta\kappa = 340 \pm 40 $ m$^{-1}$.  The colonies show a clear ordering transition, appearing as a more random, homogeneous collection of filaments below densities of 40--50 mm$^{-2}$, and displaying reticulate structures and nematic order at higher densities.

These experimental parameters correspond to a P\'eclet number $\mathrm{Pe} = (v_0 \tau \delta\kappa)^{-1} = 2.1$, with the spread of measurements from different individual filaments covering the range of $\mathrm{Pe} = 1$--$6$.  The natural lengthscale $l = v_0\tau = 1.4$ mm, with a range from individual measurements spanning 0.5--3 mm.  This leads to a dimensionless filament length of $L = 1.1$.
The filament interaction strength is difficult to quantify, but the simulation value was chosen to produce a similar average deflection as in experiments, averaged over many interactions.   Finally, the disorder-order transition of the colonies corresponds to an area coverage of $\Phi = 0.3$.  We note, however, that $\Phi$ is calculated from the physical diameter of the filaments in experiments ($d = 4\, \mu$m), whereas in the simulations an effective diameter of the interaction range ($\sigma = 7\, \mu$m) is used.  The simulations need to account for this difference in order to ensure a matching \textit{cellular} density as in the experiments, and simulation values of $\Phi = 1$ were chosen to be comfortably above the experimental disorder-order transition density.

\section{Results}
\subsection{Dynamics and transient}

\begin{figure}
    \centering
    \includegraphics{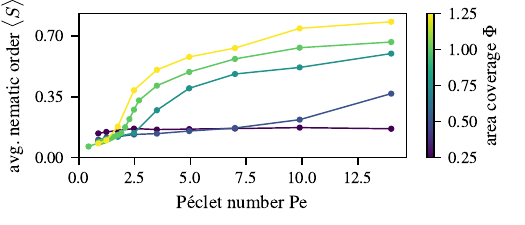}
    \caption{Continuous transition to reticulate pattern. The dependence of the average nematic order parameter $\langle S \rangle$ on \Peclet{} number is calculated as an average of $1\times 1$ sized blocks at different area coverages $\Phi$. The filament length is kept constant at $L=1$. With increasing $\Phi$ the transition from disorder to order happens at lower \Peclet{} numbers $\mathrm{Pe}$. Error bars representing the standard error are smaller than symbol size.}
    \label{fig:order_peclet_density}
\end{figure}

\begin{figure*}
    \centering
    \includegraphics{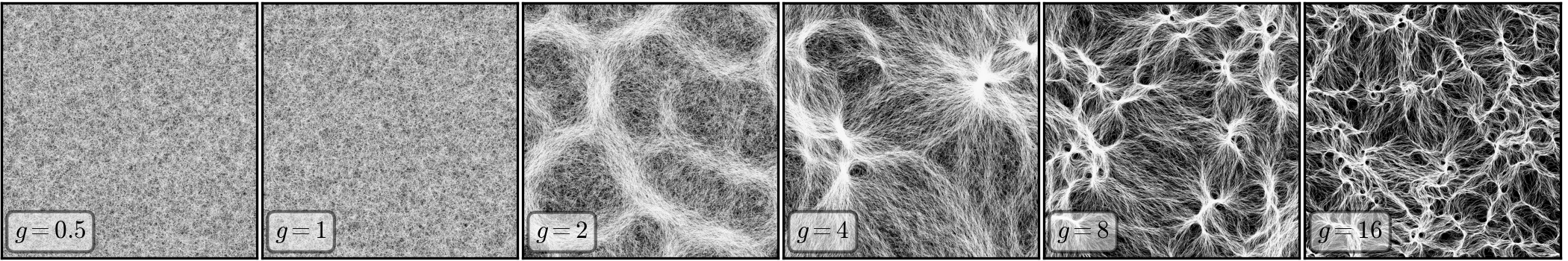}
    \caption{Effect of varying the interaction strength $g$. Snapshots of the steady-state simulated system for varying dimensionless interaction strength $g$ (see legends) at fixed $\mathrm{Pe}=3.5$, $L=1$, and $\Phi=1$. Upon increasing $g$, the colony transitions from an isotropic distribution of filaments (low $g$ limit) to a reticulate pattern ($g\approx 2$), to aster-like configurations ($g\approx 4$), and finally to strong coils ($g\gtrsim 8$).}
    \label{fig:interaction_strength}
\end{figure*}

We first investigate the dynamical aspects of the collective self-organization of cyanobacteria filaments. 
At sufficiently large number density $\rho$ and \Peclet{} number $\mathrm{Pe}$ trichomes self-organize into reticulate patterns \cite{Sumner1997, faluweki2023} as shown in Fig.~\ref{fig:progress}.  We study here the process by which this reticulate pattern emerges from an isotropic initial condition, and compare model predictions of the transient dynamics with experimental data.  In simulations, we prepare a system of filaments at area coverage $\Phi=1$ with a homogeneous distribution of filament positions and isotropic distribution of their orientations, see Fig.~\ref{fig:progress}.  For the corresponding experiments, images of the colony dynamics were cropped and contrast-enhanced to show the central 17$\times$17 mm$^2$ of the well plate, and are given in Fig.~\ref{fig:progress}.

In both experiments and simulations, the aligning interactions lead to the formation of short, dense bundles of filaments. Over time, these bundles connect and self-organize into a network topology.   During an initial transient period, the typical separation length between bundles, which provides a characteristic length scale for the patterns, slowly increases.  After a few hours in the experiments, or simulation times of $t \gtrsim 50$, the pattern then converges into a steady-state configuration of reticulate bundles of high local density, with less dense regions in between. The comparison of the transient dynamics between experiments and simulations reveals that our model quantitatively reproduces the emergence of the reticulate pattern in time; in fact, for $\tau=8$ min, as measured in Ref.~\cite{faluweki2023} a simulated time $t=75$ corresponds to 10 hours, which allows for a direct, one-to-one comparison to the experimental time scale, as in Fig.~\ref{fig:progress}. 

To characterize the dynamics of the filaments, we next consider their translational mean square displacement (MSD), defined as
\begin{equation}\label{eq:msd}
    \langle (\Delta \mathbf{x})^2 \rangle = \left\langle \frac{1}{N_t} \sum_{i=1}^{N_t} |\mathbf{x}_i(t_0+\Delta t)-\mathbf{x}_i(t_0)|^2\right\rangle\,,
\end{equation}
where the angular brackets indicate an average over initial times $t_0$.  
The MSD for a polymeric filament can in principle be computed for any part of the filaments, but 
here, for simplicity, we only use the positions of the filament heads, $\mathbf{x}_i(t)$. 
Figure~\ref{fig:msd}(a) shows the dependence of the MSD on the observation interval $\Delta t$.  Over short periods there is a ballistic regime, where the MSD is dominated by the filaments' self-propulsion and grows as $\langle (\Delta \mathbf{x})^2 \rangle \approx v_0^2(\Delta t)^2$. On longer timescales, the MSD exhibits a crossover to a diffusive regime (akin to an active Brownian particle), where $\langle (\Delta \mathbf{x})^2 \rangle =4D_\mathrm{eff}\,\Delta t$, for an effective translational diffusion constant $D_\mathrm{eff}$. 
The inset to Fig.~\ref{fig:msd}(a) shows that our simulations have comfortably reached this diffusive regime, since $\langle (\Delta \mathbf{x})^2 \rangle /4\Delta t \to \mathrm{const}$. 
Figure~\ref{fig:msd}(a) also shows the dependence of the MSD on the \Peclet{} number $\mathrm{Pe}$. As $\mathrm{Pe}$ increases, the crossover time  
between the ballistic and diffusive regimes grows as $t_\times=4D_\mathrm{eff}/v_0^2$ (see Fig.~\ref{fig:msd}(a) inset).

The temporal dynamics of the angle are complex. However, because the MSD of an Ornstein--Uhlenbeck process is known \cite{caprini2021inertial}, we can identify the long-time rotational diffusion constant as $D_R\simeq D_\omega \tau^2$, for $t\gg\tau$. 
A self-propelled particle in the absence of translational diffusion and with constant speed, whose orientation is characterized by a diffusion constant $D_R$ will exhibit a long-time diffusive behavior with an effective diffusion constant $D_\text{eff}=\frac{v_0^2}{2D_r}$ (in 2D) \cite{zottl2016emergent}. 
Using these observations, we can estimate that the filaments' heads will show a long-time diffusive behavior with $D_\mathrm{eff}= \frac{1}{2}v_0^2 \tau \mathrm{Pe}^2$. This prediction is in good agreement with the simulation results, shown as a dashed line in Fig.~\ref{fig:msd}(b), for large \Peclet{} numbers. At low $\mathrm{Pe}$, we observe some deviation from the predicted scaling, due to the fact that the estimated $D_R$ emerges from the long-time average of the curvature fluctuations quantified by $D_\omega$. At low $\mathrm{Pe}$, we have strong curvature fluctuations ($\delta\kappa\sim \mathrm{Pe}^{-1}$), thus the orientational dynamics cannot be approximated as a diffusive process, because the correlation time $\tau$ cannot be ignored in this limit.

Figure~\ref{fig:msd}(b) also shows that the ordering of filaments into bundles has a small but noticeable effect on their mobility, since switching off the interactions among filaments, i.e.\ setting $g=0$ (which removes
the possibility to form bundles) causes a small increase in their effective long-time translational diffusion $D_\text{eff}$.
In other words, the presence of the reticulate pattern alters the motility of the filaments in a subtle way but without disrupting it; that implies that the filaments do not remain confined within any particular bundle, or loop.

It is also worth mentioning that our model does not predict any dependence of the MSD on the length $L$ of the filaments, as shown in Fig.~\ref{fig:msd}(c).

While the filaments' gliding motility is of polar nature, the nematic nature of their interactions results in the nematic symmetry within the bundles, with filaments moving along them in either direction in equal parts, see Fig.~\ref{fig:nematic-angles}. The approximate equal distribution of orientations within bundles was also measured experimentally in Ref.~\cite{faluweki2023}. While the resulting patterns may look reminiscent of vortex lattices observed in other active systems \cite{doostmohammadi2016stabilization}, the filaments are not constrained to vortices and explore the paths formed by the network of bundles.

\subsection{Structure}

\begin{figure*}
    \centering
    \includegraphics{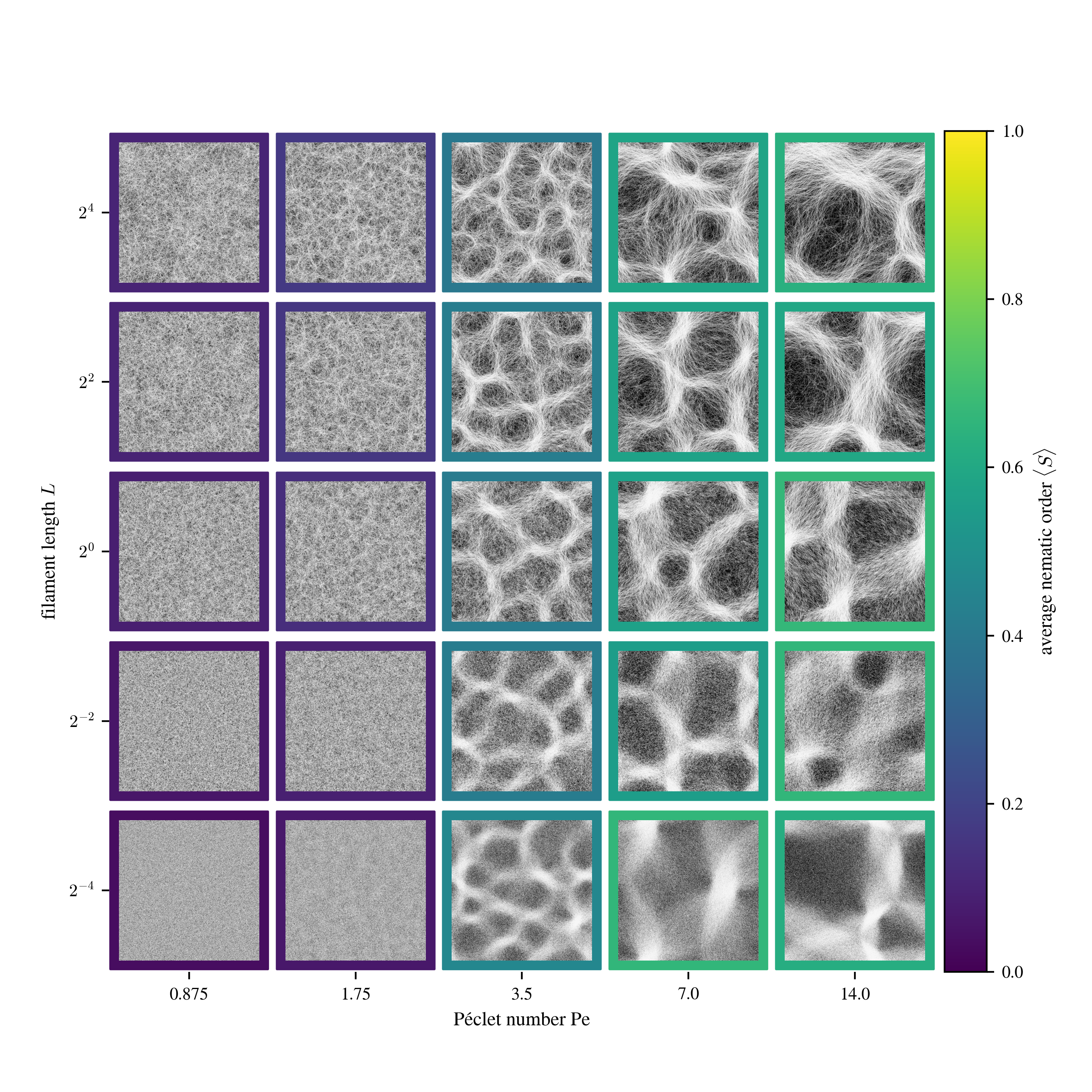}
    \caption{Emergence of reticulate patterns. The mosaic plot shows the steady-state collective behavior of the system for different filament lengths $L$ and \Peclet{} numbers $\mathrm{Pe}$. The individual panels show simulation snapshots at steady state in a $30\times30$ domain with periodic boundary conditions. The filament density is set to ensure an equal area coverage, such that $\Phi = 1$. The panels' frames indicate the system's average nematic order $\langle S \rangle$, calculated as an average of $1\times 1$ sized blocks. With increasing $\mathrm{Pe}$, the system transitions from a disordered state into one showing reticulate patterns of nematic bundles. The length scale of the emerging patterns is not affected by filament length.}
    \label{fig:combined_pe_L}
\end{figure*}

We now turn to the investigation of the structural properties of the reticulate pattern. 
Given the nematic nature of the bundles (Fig.~\ref{fig:nematic-angles}), and the fact that they can emerge from an isotropic distribution of filament orientations (Fig.~\ref{fig:progress}), we quantify the topological transition to the reticulate network topology using the 
2D nematic order parameter $\langle S\rangle$. 
To calculate $\langle S\rangle$, we sample the local orientation $\hat{\mathbf{t}}_i$ of every filament at $160L$ regularly spaced points along the filaments' length $L$, and calculate the local nematic tensor $\mathbb{Q}=\frac{1}{n} \left(\sum_i 2\mathbf{t}_i\otimes \mathbf{t}_i -  \mathbb{I}\right)$ within blocks of size $1\times 1$, where $n$ is the number of samples within a block. At this scale, the filament density is relatively homogeneous, but the blocks are large enough to provide good statistics. The local order parameter $S_\text{loc}$ is the largest eigenvalue of $\mathbb{Q}$. $\langle S \rangle$ is an average over all blocks.
Local nematic order $S_\text{loc}$ is high within bundles and lower in the dilute regions between them (see Fig.~\ref{fig:combined_pe_rho} inset at $\mathrm{Pe}=3.5$ and $\Phi=1$).

Figure~\ref{fig:combined_pe_rho} shows snapshots of the steady state of simulations at different \Peclet{} number $\mathrm{Pe}$ and area coverage $\Phi$.  The average nematic order $\langle S\rangle$ is given as a colored frame around these snapshots. 
The transition from disorder to order of our cyanobacteria colony as density increases has been explored previously \cite{faluweki2023}. Here, we observe a transition also dependent on the \Peclet{} number.
At constant $\Phi$, as $\mathrm{Pe}$ increases, the system shows a continuous transition from an isotropic disordered state to one of relatively high nematic order, driven first by the formation of bundles, and then by the emergent network-like topology (see Fig.~\ref{fig:order_peclet_density}). Upon increasing $\Phi$, the transition to a reticulate pattern occurs at lower values of $\mathrm{Pe}$, on account of the higher probability of the filaments to interact. We have previously shown \cite{faluweki2023} that the typical length scale of the reticulate pattern is directly connected to the curvature fluctuations; in our rescaled units, this predicted typical length $\ell_\mathrm{pred}=(\delta \kappa)^{-1}= \mathrm{Pe}\,v_0\tau$. Thus, as $\mathrm{Pe}$ increases, the characteristic size of the reticulate features increases. 

The second dimensionless parameter governing the filaments' dynamics is the interaction strength $g$. We explore how varying $g$ affects the structure formation in Fig.~\ref{fig:interaction_strength}.  In the limit of $g\to 0$, the system remains isotropic. Upon increasing $g$, the reticulate pattern emerges ($g\approx 2$). This is the parameter range that most closely resembles the experimental observations of young \textit{O. lutea} colonies. For larger values,  $g\approx 4$, the interactions dominate over the curvature fluctuations, and we observe aster-like structures (see e.g.\  \cite{ndlec1997self,sumino2012}, though in a different physical context), and strong coils for $g\gtrsim 8$ (see e.g.~\cite{tamulonis2014}).

We also explore the influence of filament length $L$ on the system behavior, see Fig.~\ref{fig:combined_pe_L}. Changing the filament length can affect the area coverage $\Phi$, which, as demonstrated in Fig.~\ref{fig:combined_pe_rho}, influences pattern selection.  In these simulations we therefore modulate the number density $\rho$, to keep a constant area coverage of $\Phi = 1$. Changing the filament length $L$ in this manner has no noticeable effect on the length scale of the emergent features. However, the resulting pattern is visually different. The longer filaments retain correlations in the patterns over longer times (the time it takes them to traverse their own length), leading to more fibrous-looking bundles. Shorter filaments, in contrast, produce denser bundles that show a stronger contrast in local density, when compared to inter-bundle regions; this is quantified by the more structured pair correlation function, see below. 

\begin{figure}
    \centering
    \includegraphics{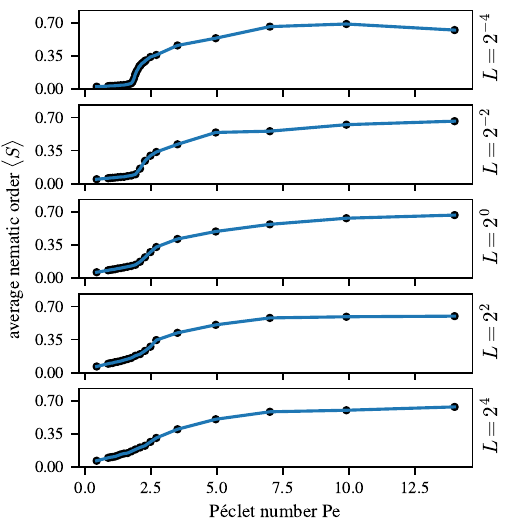}
    \caption{Approach to the continuous transition. The dependence of the average nematic order parameter $\langle S \rangle$ on \Peclet{} number is calculated as a block-average using $1\times 1$ sized blocks at different filament lengths $L$. The number of filaments is adjusted to ensure an equal area coverage of $\Phi =1$, across all panels. With decreasing $L$, the transition from disorder to order takes on the appearance of a continuous transition.} 
    \label{fig:order_peclet_length}
\end{figure}

We can obtain more precise information about the effects of the filament length $L$ in our model cyanobacteria system by considering the limiting behavior as $L$ approaches zero.   Figure~\ref{fig:order_peclet_length} shows how the average nematic order parameter $\langle S\rangle$ depends on the \Peclet{} number for different $L$. Generally, upon increasing $\mathrm{Pe}$ the system transitions from an isotropic state, characterized by a vanishingly small value of  $\langle S\rangle$, to a reticulate pattern, characterized by $\langle S\rangle\simeq 0.7$. The transition point can be identified with the inflection point of the curve at $\mathrm{Pe}=\mathrm{Pe_c}\simeq 1.8$, which does not significantly change with filament length $L$. 
The spatial extent of the filaments effectively introduces memory into the system. So, while the chance of interacting with another filament remains the same, for a constant $\Phi$, the result of an aligning interaction will be retained within the system for a time $L/v_0$. This fact enables longer filaments to retain some degree of local order, even if the system is isotropic at larger scales. The nature of the observed phase transition thus changes with increasing filament length. Short filaments experience a sharper transition from disordered to ordered states  
while longer filaments experience a much smoother transition around this point. In the limit $L\to 0$, the system for $\Phi = 1$ appears to approach a continuous transition at an apparent critical point $\mathrm{Pe_c}\simeq 1.8$. The characterization of the critical scaling in the proximity of $\mathrm{Pe_c}$ is deferred to future work. 

\begin{figure*}
    \centering
    \includegraphics{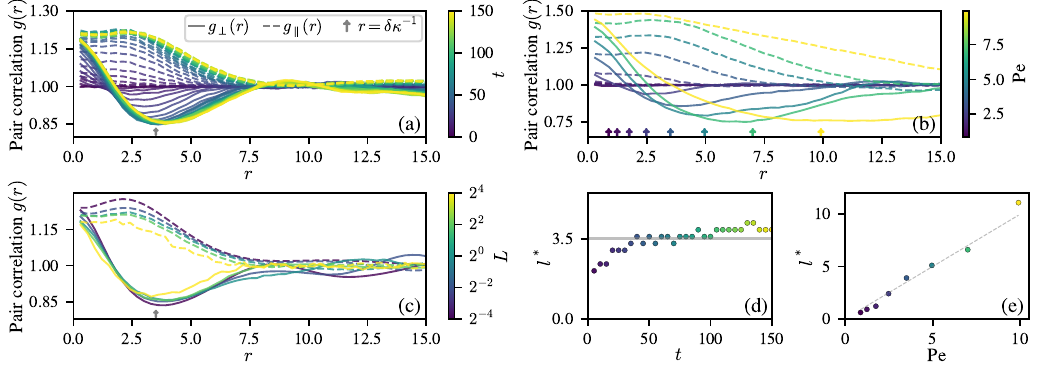}
    \caption{Emergence of structure. Filament head-head pair correlation functions (PCFs) parallel to the direction of motion, $g_\parallel(r)$ (dashed lines), and orthogonal to it, $g_\perp(r)$ (solid lines), see Eqs.~\eqref{eq:g_par}-\eqref{eq:g_perp}, quantify the structure of the reticulate pattern. (a) The time evolution (color axis) of these functions is shown for a system with area coverage $\Phi=1$, filament length $L=1$ and \Peclet{} number $\mathrm{Pe}=3.5$.  
    As time advances and the reticulate pattern forms, $g_\parallel(r)$ and $g_\perp(r)$ converge to a steady state.  
    For a system with a distribution of curvatures $\delta\kappa$, the minimum of $g_\perp(r)$ coincides with $\mathrm{Pe}=(v_0 \tau \delta\kappa)^{-1}$, indicated by an arrow on the $x$-axis. 
    (b) Steady-state PCFs are shown for different $\mathrm{Pe}$, keeping filament length $L=1$ and $\Phi = 1$. The minima in $g_\perp(r)$ coincide with the length scales predicted from the distributions of curvatures, 
    indicated by color-coded arrows. The shoulder in $g_\parallel(r)$ moves to larger distances with increasing $\mathrm{Pe}$ (except for $\mathrm{Pe}= 9.9$, where finite size effects are visible).
    (c) Effect of filament length $L$ (color bar) on the PCFs at $\mathrm{Pe}=3.5$. The number density $\rho$ is scaled here to fix $\Phi = 1$. Independently of $L$, the minimum in $g_\perp(r)$ coincides with the system's Pe, as denoted by an arrow. The magnitude of the extrema in $g_\parallel$ and $g_\perp$ 
    decrease with increasing filament length, indicating less well-defined patterns.
    (d) Temporal evolution of the typical length scale $\ell^*$ associated to the reticulate pattern, obtained from the minimum of $g_\perp(r)$ in panel (a). At long times, $\ell^* \to 3.5$, which is the value of $\mathrm{Pe}$. (e) Dependence of $\ell^*$ on $\mathrm{Pe}$ showing that the minimum of $g_\perp(r)$ matches the \Peclet{} number and is a good measure of the reticulate's typical length scale; the dashed line corresponds to the identity.}
    \label{fig:pair_corr}
\end{figure*}

A powerful method to obtain structural information in soft-matter systems is the pair correlation function (PCF) $g(r)$ \cite{kirkwood1942radial}. This is defined as the probability of finding another particle within a volume $dV$ at distance $r$ from a reference particle, relative to a homogeneous distribution. In polymer physics, it is possible to compute the PCF for any pair of monomers within the same chain or between different chains \cite{zimm1948scattering}. In our system, a simple choice is to consider the head-head PCF, where only the head bead of each filament is considered. Furthermore, given the long persistence length of the filament it is convenient to consider $g(r)$ along the directions parallel and normal to the direction of motion of the filament's head.  We thus define the head-head parallel PCF
\begin{align}
    g_\parallel(r)= & \frac{1}{2\rho N_t w \Delta r }\left \langle  \sum_{i,j} \delta\left(|(\mathbf{x}_i-\mathbf{x}_j)\cdot \mathbf{t}_i^0|-r \right) \right \rangle\,, \label{eq:g_par}  
\end{align}
and the head-head perpendicular PCF
\begin{align}
    g_\perp(r)= & \frac{1}{2\rho N_t w \Delta r }\left \langle  \sum_{i,j} \delta\left(|(\mathbf{x}_i-\mathbf{x}_j)\times \mathbf{t}_i^0|-r\right) \right \rangle\,. \label{eq:g_perp}
\end{align}
In Eq.~\eqref{eq:g_par}-\eqref{eq:g_perp}  $\delta(r)$ is the Dirac delta distribution, the sums extend to all distinct pairs of filaments $i$ and $j$, and $\mathbf{t}_i^0$ is the tangent unit vector to the head bead; we sample filaments' heads within a narrow rectangle of width $w=10\sigma$ either parallel or perpendicular to the reference filament head's orientation, and discretize the scalar distance $r$ in bins of size $\Delta r$. 

Figure~\ref{fig:pair_corr}(a)) shows the emergence of the reticulate structure over time, as quantified by $g_\parallel(r)$ and $g_\perp(r)$. As the system starts in an isotropic state (see Fig.~\ref{fig:progress}), initially  $g_\parallel(r) = g_\perp(r)=1$. As the reticulate pattern forms, $g_\parallel(r)$ develops a pronounced shoulder, and $g_\perp(r)$ develops a distinct minimum. 
The significance of these features can be related to the reticulate geometry. For example, moving forward from any particular filament's head, along the direction of its tangent vector, one will be more likely to find other heads so long as they are within a bundle of filaments. Once the bundle bends away, $g_\parallel(r)$ will decrease. This is reflected in a shoulder in $g_\parallel(r)$. Considering instead exploring in the direction normal to the path of a reference filament's head, one is also more likely to find other heads, but only so long as they are within the same bundle of filaments.  As the bundles are relatively narrow, this probability decays quickly and drops below one in the regions between bundles, but then grows back when another bundle is encountered; this arrangement produces a minimum in $g_\perp(r)$. 
The position of this minimum $\ell^* =\operatorname{arg\,min} g_\perp(r)$ serves as a measurement of the pattern's length scale;
as the pattern coarsens, the depth of the minimum in $g_\perp(r)$ and height of the shoulder in $g_\parallel(r)$ increase, indicating the growth of the pattern length scale, as was observed in Fig.~\ref{fig:progress}. The position of the minimum eventually settles on a value close to that of $\ell_\mathrm{pred}= \mathrm{Pe}\,v_0\tau$.

Upon increasing $\mathrm{Pe}$, the position and magnitude of the shoulder in $g_\parallel(r)$ and the minimum in $g_\perp(r)$ grow, see Fig.~\ref{fig:pair_corr}(b). Simulations performed at the highest \Peclet{} numbers show some finite size effects, as the scale of the pattern becomes comparable to the simulation domain. 
The prediction that the reticulate pattern's length scale is determined by the width of the curvature distribution $\delta \kappa$ \cite{faluweki2023} holds throughout these simulations, as shown by the position of the shoulder of $g_\parallel(r)$ and the minimum of $g_\perp(r)$. The filament length $L$ has seemingly no influence on the typical scale of the emerging pattern. However, ordered and disordered regions are more clearly separated for shorter filaments (Fig.~\ref{fig:pair_corr}(c)), as evident in the more pronounced structure of both parallel and perpendicular PCFs.

We stress that, despite the fact that our system is topologically non-trivial and thus violates the condition of homogeneity used to derive the PCF \cite{hansen2013theory}, the latter does reflect important structural information about the filament colony.

Figure~\ref{fig:pair_corr}(d)-(e) show the dependence of $\ell^*$ on time and \Peclet{} number, respectively, as based on the results in Fig.~\ref{fig:pair_corr}(a)-(b). The position of the minimum of $g_\perp(r)$ converges to the \Peclet{} number of the simulation, which in our rescaled units, corresponds to the prediction for the length scale of the pattern $\ell_\mathrm{pred}=\mathrm{Pe}\,v_0\tau$ \cite{faluweki2023}.

\section{Conclusions}

We have studied the dynamics and structural properties of colonies of filamentous cyanobacteria as they undergo a topological transition to a reticulate pattern. The ability of cyanobacteria to glide over one another while experiencing weak aligning interactions enables the formation of a disordered network-like structure of filaments within bundles that link together.
In addition to the network topology, confined cyanobacteria systems can form topological defects \cite{gong2023,repula2024}, bearing a striking resemblance to some active nematic systems \cite{maryshev2020pattern,kruger2023hierarchical}. Other than these confined systems, where defects can only resolve the mismatch of order by non-local reconfiguration, the unconfined cyanobacteria can glide over one another, resulting in dynamics not dominated by topological defects.

Filamentous cyanobacteria, which we have termed \textit{active spaghetti} \cite{faluweki2023}, constitute a new class of active matter, due to the path-tracking nature of their gliding motility. We derived the nondimensional equations for a minimal model of their behavior that is firmly grounded in experimental evidence. The model captures the transient dynamics leading to the formation of the reticulate pattern.  It also predicts a crossover in the mean-square displacement of filaments between ballistic and diffusive regimes, which can be rationalized in terms of an effective diffusivity. We explored the dynamics for different filament lengths, and found that the order parameter appears to approach a continuous phase transition in the limit of very short filaments.
This is in contrast to the well-studied Vicsek model \cite{vicsek1995} which exhibits a weak first-order transition \cite{gregoire2004onset}. Furthermore, our system does not exhibit an ordered phase where all filaments point in the same direction, on account of the curvature fluctuations present in our model.
The \Peclet{} number dictates the typical length scale of the patterns, and together with the area coverage $\Phi$ describes the relevant fields of the nonequilibrium phase diagram for our system. Inspired by polymer physics, we introduced parallel and perpendicular pair correlation functions, which capture the fundamental structural properties of the reticulate bundles. In particular, the minimum of $g_\perp(r)$ faithfully predicts the typical length scale  $\ell^*$ of the pattern. These observations can help further understand the formation and properties of cyanobacterial biomats.

\begin{figure}
    \centering
    \includegraphics[width=\columnwidth]{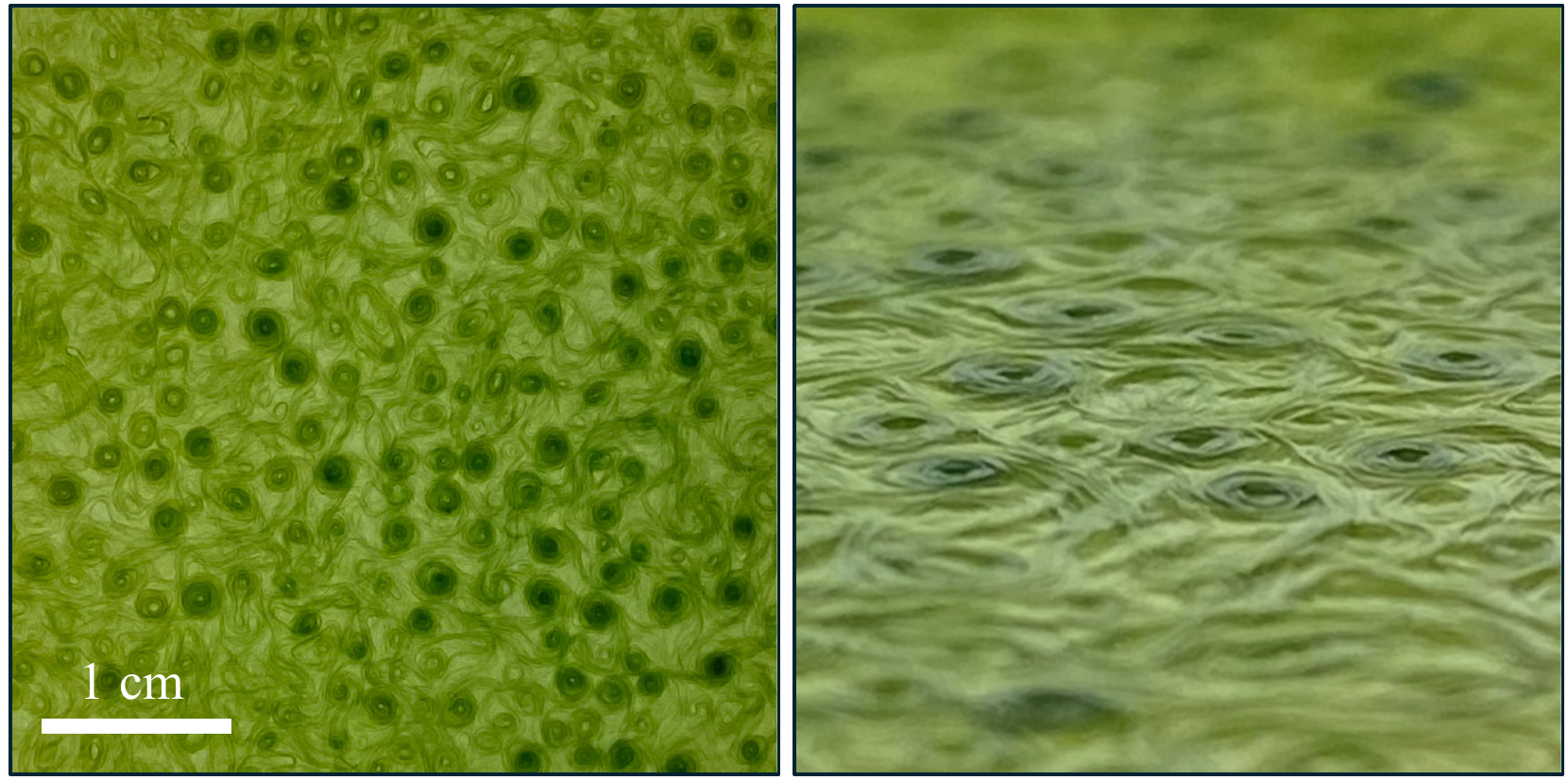}
    \caption{Mature cyanobacteria biofilm. \textit{O. lutea} grown on agar for six months shows a more complex, three-dimensional structure.  Photographs of the top surface of the biomat (left panel: face-on view; right panel: low-angle shot of the same sample) show a complex topography of raised, filled circular features, which develop out of an earlier template of a more open, reticulate structure.}
    \label{fig:agar}
\end{figure}

Biofilms and biomats are self-organized microbial communities that differentiate and develop phenotypic adaptation both in space and time \cite{stoodley2002biofilms}. It stands to reason that the form of the reticulate patterns studied here has a biological function sculpted by survival needs. The reticulate network then serves as a foundation for mature biofilms that adapts to the features of the surrounding environment.  Even growing as films on glass culture bottles, these biofilms can easily reach 100 $\mu$m thick, much thicker than an individual filament. Because gliding cyanobacteria are surface associated, a natural question is then to inquire how the surface properties influence the growth and maturation of the biofilm \cite{krsmanovic2021hydrodynamics,snowdon2023surface}. 
As an initial exploration of the effects of substrate stiffness on colony formation, we also cultivated \textit{O. lutea} on agar plates, see Fig.~\ref{fig:agar}.  These were prepared by mixing 1.5\% agar, by weight, with a 1:50 dilution of BG11 broth in deionized water, following the protocols of Ref.~\cite{SAG}.  Every eight weeks, 2 ml of 1:100 BG11 broth were added to the plates.  Samples were otherwise left undisturbed, growing under identical conditions to those described for the transient experiment (Fig.~\ref{fig:progress}), and following Ref.~\cite{faluweki2023}.  Over time, the initially reticulate pattern developed into a three-dimensional structure, with the open areas of the network filling in and growing upward, which was templated on the original network boundaries. These observations point at the physical richness of this biological system and invite future exploration.

\section{Acknowledgments}

We thank Maike Lorenz (SAG G\"ottingen) for support with cyanobacteria cultures, Stefan Karpitschka from the Max Planck Institute for Dynamics and Self-organisation (MPIDS) and Jack Paget (Loughborough University) for discussions, and Graham Hickman at Nottingham Trent University (NTU) for microscopy support. We gratefully acknowledge Ellen Nisbet (University of Nottingham) for helpful discussions, and Alex Luke (NTU) for preliminary work exploring the imaging of biofilms. 
Microscopy facilities were provided by the Imaging Suite at the School of Science and Technology at NTU. Numerical calculations were performed using the Sulis Tier 2 HPC Platform funded by EPSRC Grant EP/T022108/1 and the HPC Midlands+ consortium. We gratefully acknowledge use of the Lovelace HPC service at Loughborough University. M.K.F. was partly sponsored by the Malawi University of Science and Technology. This work was supported by the MPIDS. J.C. was supported by EPSRC grant EP/W522569/1 and  UKRI/Wellcome grant EP/T022000/1–PoLNET3. N.D. was supported by a BBSRC doctoral training programme (BB/T0083690/1). 

\section{Author contributions}
J.C. developed the simulation model and performed the simulations. M.K.F. and N.D. performed experiments. L.G. designed experiments. J.C., L.G., and M.G.M. designed the research. 
J.C., L.G., and M.G.M. wrote the manuscript with inputs from all authors.

\section{Competing interests}
The authors declare no competing interests.

\section{Data availability}
All data needed to evaluate the conclusions in the paper are present in the paper.
All micrographs used are shown in the paper. Micrographs in their original resolution are available upon reasonable request to L.G.

\section{Code availability}
The simulation code and output data are available upon reasonable request to M.G.M.

\bibliography{cyano}

\end{document}